\documentstyle[aps]{revtex}
\begin{document}
\input epsf
\draft
\twocolumn[\hsize\textwidth\columnwidth\hsize\csname
@twocolumnfalse\endcsname
\preprint{PURD-TH-98-07,  UH-IfA-98-30,
 SU-ITP-98-32,  hep-ph/9805209}
\date{May 1998}
\title{Cosmic strings from  preheating}
\author{I. Tkachev,$^{1,2}$ S. Khlebnikov,$^1$  L. Kofman,$^3$ and A.
Linde$^4$}
\address{
${}^1$Department of Physics, Purdue University, West Lafayette, IN
47907, USA \\
${}^2$Institute for Nuclear Research of the Academy of Sciences of
Russia, Moscow 117312, Russia \\
${}^3$Institute for Astronomy, University of Hawaii, 2680 Woodlawn Dr.,
Honolulu, HI 96822, USA \\
${}^4$Department of Physics, Stanford University, Stanford, CA
94305-4060, USA}
\maketitle
\begin{abstract}
We investigate nonthermal phase transitions that may occur after
post-inflationary preheating in a simple model of a two-component scalar 
field with the
effective potential
$\lambda (\phi_i^2 - {\rm v}^2 )^2/4$, where $\phi_1$ is
identified
with the inflaton field. We use three-dimensional lattice simulations to
investigate  the full nonlinear dynamics of the model. Fluctuations of the
fields generated during and after preheating temporarily make the 
effective
potential
convex in the $\phi_1$ direction. The subsequent  nonthermal phase
transition
with symmetry breaking leads to formation of cosmic strings even for ${\rm v} \gg 10^{16}$ GeV.
This mechanism of string formation, in a {\em modulated} (by the
oscillating field $\phi_{1}$) phase transition, is
different from the usual Kibble mechanism.
\end{abstract}
\pacs{PACS: 98.80.Cq\hskip 1.1cm PURD-TH-98-07,~ UH-IfA-98-30,~
SU-ITP-98-32\hskip 1.1cm hep-ph/9805209}
\pacs{\hskip 2 cm {\it This paper is dedicated to the memory of David Abramovich Kirzhnits}}\vskip2pc]

 The theory of cosmological phase transitions proposed in 1972 by David Kirzhnits gradually became one of the most  essential parts
of modern cosmology   \cite{Kirzhnits}. Originally it was assumed that such phase transitions occur in a state of thermal equilibrium when the temperature decreased in the expanding universe. Recently it was found that  large fluctuations of scalar and vector fields produced during preheating after
inflation \cite{KLS1} may lead to specific nonthermal   phase
transitions which occur far away from the state of thermal equilibrium \cite{ptr}.  To investigate  these phase transitions one should
study the self-consistent nonlinear dynamics of quantum fluctuations  amplified by  parametric resonance. This
is a very complicated task. Fortunately, fluctuations
of Bose fields generated during preheating have very large
occupation numbers and
can be considered as interacting classical waves, which allows
one to perform a full  study of all nonlinear effects during and after
preheating using lattice calculations \cite{KhTk1}.
These calculations, as well as analytical estimates  \cite{KLS1,ptr,KhTk1,KhTk2,KhTk3,KhTk4,Prokopec}
have already shown that fluctuations can grow large enough for
cosmologically interesting phase transitions to occur.  Lattice calculations can be used to directly simulate nonthermal
phase transitions and formation of topological defects.
This  made it possible  to go
beyond the early attempts to study such phase transitions numerically
\cite{BoyKaw}, which neglected crucial
backreaction effects beyond the Hartree approximation.

We made  a series of lattice simulations of nonthermal
phase transitions, focusing primarily on the possibility of
generation of topological defects \cite{TKKLS,kklt}.
 The model with one scalar field
with the potential $\lambda (\phi^2 - {\rm v}^2 )^2 /4$
 has a broken discrete symmetry $\phi \rightarrow - \phi$.
In this model we have found formation of domain
structure  \cite{TKKLS}. If the inflaton field $\phi$ strongly couples to
another scalar field $\chi$, the phase transition is strongly first
order; we observed formation of bubbles of the new phase \cite{kklt}.

 In this Letter we will study string formation in the theory of a
two-component
scalar field $\phi_i$    with the effective potential $\lambda
(\phi^2 - {\rm
v}^2 )^2 /4$, where  $\phi^2= \sum\phi_i^2$. This model has $O(2)$
rotational
symmetry and allows
for string  formation  \cite{TKKLS} (for early conference
reports see
\cite{tokyo}).
Recently Kasuya and Kawasaki  reported formation of defects 
in 2d lattice simulations for  ${\rm v} \lesssim 3  \times 10^{16}$ GeV
 \cite{KawNew},  which confirmed  our general
conclusion
concerning  the phase transition in this model
\cite{TKKLS,tokyo}.
 However,  scattering of
particles, as well as the nature of topological defects in two
dimensions and
in three dimensions are quite different. In particular, there are no
strings in
this model in two dimensions. We have found that in the realistic case of
three dimensions the
 generation of fluctuations is much more effective, and
 string production is possible even if  ${\rm v}$ is as large as $6\times  
10^{16}$ GeV.

In the model with the two-component field $\phi_i$ we
can always
rotate fields in such a way that initially $\phi_2 = 0$, and the
field $\phi_1$
plays the role of the  classical oscillating inflaton field. We denote by
$\phi(0)$ the value of the field $\phi_1$   at the  moment when
inflation ends and the inflaton field begins to oscillate.
 It is convenient to work with the rescaled conformal time $\tau$, where
$\sqrt{\lambda}\phi(0) dt= a(\tau)d\tau$, $a(0)=1$, and
perform conformal transformation of the fields, $\varphi = \phi(\tau)
a(\tau)/\phi(0)$.
We also will use the rescaled spatial coordinates ${\bf x} \to
\sqrt{\lambda}\phi(0){\bf x}$.
The initial conditions at the beginning of preheating are
determined by the preceding stage of inflation.
We define the beginning of preheating as the moment  $\tau=0$  when the
velocity of the field $\phi$ in conformal time is zero.
 This happens when
 $\phi(0)\approx 0.35 M_{\rm Pl}$ and
$a(\tau) \approx 0.51\tau+1$ \cite{KhTk1}.

In the new variables, the equations of motion become
\begin{equation}
{\ddot \varphi_i} - \nabla^2 \varphi_i  + (\varphi^2
-v^2a^2)\varphi_i= 0 \; ,
\label{eqm}
\end{equation}
where ${v}={\rm v}/\phi (0)$.
Equation (\ref{eqm}) contains only one  parameter, ${
v}$. The coupling constant $\lambda$ is hidden in the initial conditions
for fluctuations; these were chosen as described in Ref. \cite{KhTk1}.

The full nonlinear equations of motion
(\ref{eqm}) were solved numerically directly in the configuration
space.
The computations were done on $128^3$ lattices
with the box size $L=16 \pi$,  with the expansion
of the
universe assumed to be radiation dominated.
There are several important quantities that can be measured
in the simulations during the
evolution of the system. One can define the zero mode
$\varphi_{0i} = \langle\varphi_i(x,\tau)\rangle$ for  each of the
components $i=1,2$ of the field $\varphi$, and variances $\langle (\delta
\varphi_i)^2\rangle$, which measure
the average magnitude of fluctuations $\delta \varphi_i =
 \varphi_i(x,\tau)-\varphi_{0i}$.
Since the system is homogeneous on large scales, averages in these
relations
can be understood as volume averages. This is equivalent to taking an
average over realizations of
the initial data (the ensemble average). The averaged
quantities depend
only upon time and do not depend upon spatial coordinates ${\bf x}$.

\begin{figure}[Fig6]
\centering
\leavevmode\epsfysize=5.5cm \epsfbox{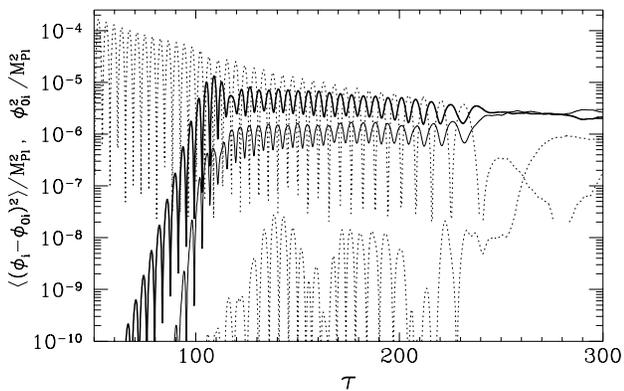}\\
\

\caption[Fig6]{\label{f2m_strings} Variances of the field components
$\phi_2$
and $\phi_1$ (upper  and lower
solid curves respectively) as well as  the zero modes of these fields
(upper dotted curve is $\langle \phi_1\rangle $ and lower dotted curve is
$\langle \phi_2\rangle$) are shown as functions of conformal time. }
\end{figure}

We performed calculations for various ${\rm v}$ and $\lambda$. In the beginning we present
results for ${\rm v}
= 3\times
10^{16}$ GeV and   $\lambda =10^{-12}$.  In the last part of the paper we also present result for different values of ${\rm v}$ and for $\lambda = 10^{-13}$. The behavior
of the zero modes
$\phi_{0i} = \langle\phi_i\rangle$,  and of the variances,
$ {\langle (\phi_i-\phi_{0i})^2 \rangle}$, for each of the field
components is
shown in Fig.~\ref{f2m_strings}.

Let us explain the evolution of the fluctuations.
For such a small value of ${\rm v}$,
preheating   is completed at $\tau \approx 100$, well before the phase 
transition with symmetry breaking, which  occurs
at $\tau \approx 240$.
During the  stage of parametric
resonance $\tau \leq 100$, the term  $v^2a^2$ in (\ref{eqm})  can be 
neglected. The equations for the mode functions of fluctuations
in the  directions of  $\phi_1$ and $\phi_2$  are
\begin{equation}
 \ddot{\varphi}_{ik} +(k^2+q_i\varphi_{01}^2)\varphi_{ik}  = 0 \, ,
\label{fluc}
\end{equation}
where the background inflaton oscillations $\varphi_{01}(\tau)$
 are given by an elliptic function.
The  resonance parameter is different for different
components: $q_1=3$ and $q_2=1$.
For both components there is a single instability band, but the locations
of the bands and  the strengths of the resonance are 
different~\cite{GKLS}.
The resonance in the ``inflaton'' direction $\varphi_1$ is weak,
the maximal value
of the characteristic exponent of the fluctuations
$\varphi_{1k} \propto e^{\mu_1 \tau}$ is $\mu_1 \approx 0.036$; the
resonance
in the perpendicular  direction $\varphi_2$ is much   stronger
and broader, $\mu_2\approx 0.147$.

However,  the actual growth rate of the fluctuations $\varphi_{1k}$ 
is not smaller but larger than that of the
$\varphi_{2k}$   fluctuations. Indeed, the cross-interaction term
$\varphi^2_{1}\varphi^2_{2}$  leads to the production of
 $\delta \varphi_1$ fluctuations in the process of rescattering
\cite{KhTk3}.
As a result,   the amplitude $\varphi_{1,k}$
grows   as $\varphi_{1k} \propto e^{2 \mu_2 \tau}$,
 where $2\mu_2=0.294 $ \cite{tokyo},
as clearly seen in Fig.~\ref{f2m_strings}.
This growth is much faster than in the 2d lattice simulations of Ref.
\cite{KawNew}.

We see that during the time
interval between completion of preheating and
the phase transition,
$100 \alt \tau \alt 240$,
the zero mode of $\phi_1$
decreases faster than the variances (both are decreasing due to
the expansion of the universe, but in addition the zero mode
continues to decay into field fluctuations), and
by the time $\tau \approx 240$ the field variance is as big as its
zero mode.

\begin{figure}[Fig8]
\centering
\leavevmode\epsfysize=5.5cm \epsfbox{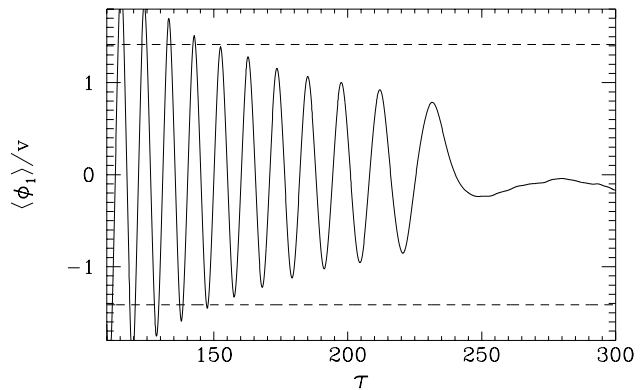}\\
\

\caption[Fig8]{\label{zmod_string} Time dependence of the zero mode of
$\phi_1$.}
\end{figure}

Time interval $100 \alt \tau \alt 240$ is dominated by  rescattering
of all modes
\cite{KhTk1}.  During this time interval, the symmetry (at least
in one direction) is restored by
large fluctuations.
To see this, we
 study the time dependence of the zero mode of $\phi_1$,
which is
shown in Fig.~\ref{zmod_string}. We immediately see some peculiarity.
In the potential of the form $(\phi^2-{\rm v}^2)^2$ the field cannot
oscillate
near $\phi=0$ with amplitude smaller than $\phi_c \equiv\sqrt{2} {\rm v}$.
This critical value of the amplitude is shown in Fig.~\ref{zmod_string} by
the dashed line, and we see  that the field does oscillate with an
amplitude
smaller than $\phi_c$. At some point, the amplitude of the oscillations 
even becomes smaller than ${\rm v}$, i.e. the field is
oscillating on the top of the tree potential without rolling down to its
minimum. This means that the tree potential is significantly
altered by the
interaction with the background of created fluctuations, and the
symmetry is
restored.

\begin{figure}[Fig9]
\centering
\leavevmode\epsfysize=5.5cm \epsfbox{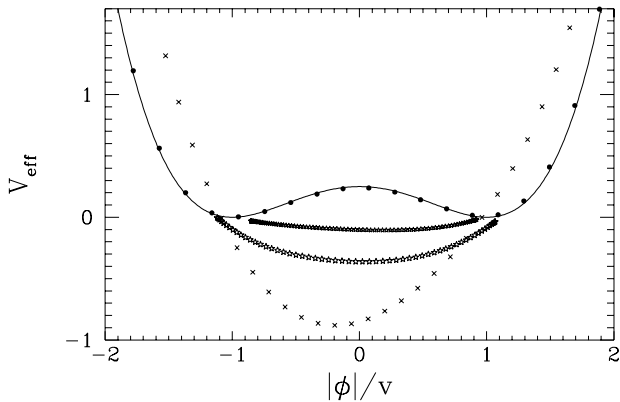}\\
\

\caption[Fig9]{\label{eff_pot_string} Reconstruction of the
effective potential in $\phi_{1}$ direction
at several moments
of time. Dots correspond to $\tau \approx 90$, diagonal crosses to
$\tau \approx 110$, larger stars to $\tau \approx 180$ and smaller
stars to
$\tau \approx 215$. Solid line is the tree potential.}
\end{figure}

We can reconstruct the effective potential
 $V_{\rm eff}(\varphi)$ using the
already calculated
function
$\varphi_{0i}(\tau )=\langle\varphi_i\rangle $ and the definition
\begin{equation}
{\ddot \varphi_{0i}}  + dV_{\rm eff}/d\varphi_{0i}= 0  \; .
\label{eff_pot_def}
\end{equation}
This definition of the effective potential is perfectly legitimate
in the case
when corrections to terms with the time derivative are small, which
appears to be the case. (Expansion of the universe will be properly
included
when
we  work in
the conformal coordinates, as in Eq. (\ref{eff_pot_def}), and then
``rotate''
potential back to the synchronous frame and to the original fields.)
 This method allows one to find $V_{\rm eff}$ up to a constant.
The reconstruction along $\varphi_{1}$ direction
is shown for several moments of time in Fig.~\ref{eff_pot_string}.
At $\tau \approx 90$ the effective potential, shown by dots,
coincides to a
very good accuracy with the tree potential, shown by the solid line.
This moment is not far away from the end of the exponential growth of
fluctuations, which  
occurs at
$\tau \approx 100$, but fluctuations at $\tau \approx 90$ are still rather
small. At $\tau
\approx 110$ fluctuations are large, and
the effective potential is completely different: it has
only one
minimum.
Note that the potential is slightly asymmetric with respect to the
change of
sign of the field $\phi_1$. This happens because in our simulations we are
sampling the
effective
potential in the time-dependent background, when the points with  
positive and
negative $\phi$ correspond to different moments of time. The  
``instantaneous"
potential would be exactly symmetric, with a minimum at $\phi_1=0$, and symmetry is restored, at least in the $\phi_{1}$ direction.

Still, the  existence of an oscillating zero mode $\phi_1$ implies that   symmetry between   states with $\phi_1 > 0$ and $\phi_1 < 0$ is not exact. Thus one wonder whether one can say that symmetry is restored if the effective potential has the minimum at $\phi_1 = 0$, or one should not say so until the amplitude of oscillations of the field $\phi_1$ completely vanishes.

In our opinion, if the effective potential has a minimum at $\phi_1 = 0$, this implies that there is no {\it spontaneous} symmetry breaking, so in this sense the symmetry $\phi_1 \to - \phi_1$ is restored.  From a more pragmatic point of view, the issue of symmetry restoration  and subsequent symmetry breaking is important mainly because it is related to production of topological defects. Therefore instead of debating which definition of symmetry restoration is better, we will  try to find whether the topological defects are produced. Indeed we will see that cosmic strings are produced as a consequence of preheating.

At $\tau \approx 240$ the field fluctuations become diluted by the
expansion
of the universe to the extent sufficient for symmetry breaking.
This event can be seen both in Fig.~\ref{f2m_strings} and in
Fig.~\ref{zmod_string}. At that time, the zero 
mode nearly vanishes. 
This happens
not because of symmetry restoration, but because the field rolls
down to the
$O(2)$ symmetric valley of minima of the effective potential  in all
possible
directions in the field space, which, after averaging, gives $\langle
\phi_1\rangle \ll {\rm v}$.

\begin{figure}[Fig8]
\centering
\leavevmode\epsfysize=4cm \epsfbox{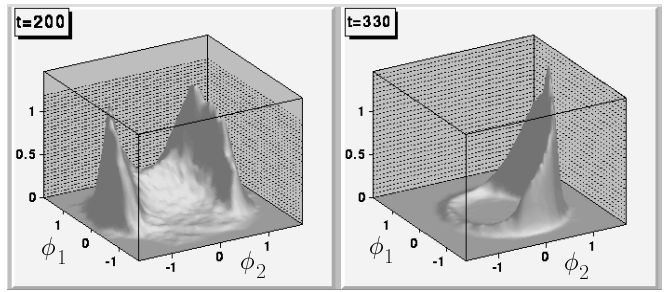}\\

\

\caption[Fig8]{\label{pdf_string} The joint probability distribution
function of $\phi_1$ and $\phi_2$ before and after the phase
transition. Fields $\phi_1$ and $\phi_2$ are shown in units of ${\rm v}$.}
\end{figure}

A useful quantity to consider   is  the probability
distribution function   $P(\phi_1,\phi_2, t)$  shown in
Fig.~\ref{pdf_string}.
The first of these two figures
shows   that
at $\tau \sim 200$
  the maximum of the probability distribution  oscillates near
$\phi_1 = 0$ in
the $\phi_1$ direction. This distribution has only one maximum in
the $\phi_1$
direction. When this maximum approaches $\phi_1 = 0$, the distribution
$P(\phi_1,\phi_2, t)$ is approximately symmetric with respect to the
change
$\phi_1 \to -\phi_1$. On the other hand,    there are two maxima of the
probability distribution with respect to the field $\phi_2$. They are
concentrated   near $\phi_2  \approx \pm {\rm v}$, which means that the
symmetry $\phi_{2} \to - \phi_{2}$  is broken.

This   implies that the universe at that time becomes divided into domains
filled with the field $\phi_2 \approx \pm {\rm v}$. These domains are
separated by
two-dimensional domain walls, the surfaces where $\phi_2 = 0$. When the
distribution of the scalar field $\phi_1$ oscillates, the center
of this
distribution moves, but if it is wide enough, there always will be
two-dimensional surfaces where $\phi_1 = 0$. Intersection of
these surfaces
with the domain walls $\phi_2 = 0$ form strings, on which $ \phi_1 =
\phi_2 = 0$. One can easily find out that when one moves around such a
string, the phase $\alpha =   \arccos  (\phi_1/|\phi |)$ changes by
$2\pi$.
Thus these strings are topologically stable. Most of them form
string loops
which move (expand, shrink, and expand again) when the distribution of the
field $\phi_1$ oscillates. This mechanism of string formation  is
different
from the Kibble
mechanism.

 Gradually the amplitude of  fluctuations of the fields $\phi_i$
decreases, and symmetry $\phi_1 \to -\phi_1$ also breaks down.
The choice of  direction of the symmetry breaking will depend on the shape of the effective potential, but also   on the amplitude and phase of the oscillations of the zero mode $\langle \phi_1 \rangle$, and on the width of the probability distribution $P(\phi_1,\phi_2, t)$. If the
width of the probability distribution $P(\phi_1,\phi_2, t)$ in the
$\phi_1$ direction is sufficiently large, then this  phase transition modulated by the oscillations of the field $\phi_1$ may lead to
formation of long strings, which may have interesting cosmological
consequences.

After the modulated phase transition, which occurs at $\tau =240$, we see an 
$O(2)$ symmetric ring in the probability
distribution $P(\phi_1,\phi_2, t)$, superimposed
with a peak in some random direction (which does not coincide
either with $\phi_{1}$ or with $\phi_{2}$), see the second figure in
Fig.~\ref{pdf_string}. The existence of
the ring
shows that the absolute value of the field after spontaneous
symmetry breaking
is close to ${\rm v}$, and that in different points of space the vector
$(\phi_1,\phi_2)$ looks in all possible directions, which
indicates the
presence of strings.  The peak along the random
direction represents additional {\it spontaneous} symmetry breaking, which
appears because
it is energetically preferable for all vectors
$(\phi_1,\phi_2)$ to look
in the same direction. With time, the ring will disappear, and the width 
of the peak will decrease.

\begin{figure}[Fig9]
\centering
\leavevmode\epsfysize=8cm \epsfbox{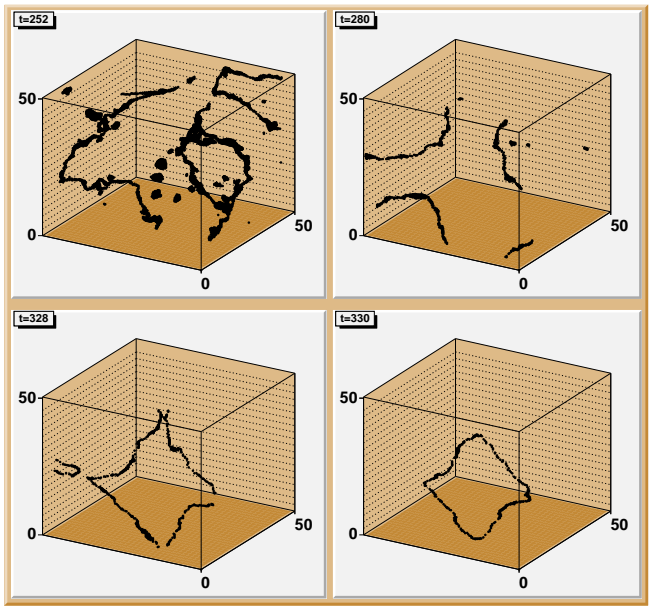}\\
\

\caption[Fig1]{\label{strings_strings} The process of string formation for ${\rm v} = 3\times 10^{16}$ GeV and $\lambda = 10^{-12}$.}
\end{figure}

Strings can be detected directly. For that purpose,
we had plotted 3D coordinates of the grid points where $|\phi |$ is close
to zero. A series of such images describing several
different stages of this process is shown in Fig.~\ref{strings_strings}.
Strings are clearly seen, and we can observe the formation of
one large loop.

At times $100< \tau <240$,
while the zero mode was still oscillating, we had observed  abundant
formation and subsequent annihilation of string loops.
Many strings appear when the oscillating zero mode 
$\langle\phi_1(t)\rangle$ passes through
$\phi_1 = 0$. Then they disappear when $\langle\phi_1\rangle$ grows and
appear when it becomes small again.
Fig.~\ref{strings_strings}, on the other hand, shows the behavior of 
strings 
after the phase transition,  when  the field $\phi_1$ is not  capable
of rolling
over the top of the effective potential. At this stage the new
strings are not
produced, and the old
ones move slowly, changing their position because of the string tension.

 It is interesting that at time $\tau =280$ there is only one string 
loop
in the integration box (upon account of the periodic boundary 
conditions),
see Fig.~\ref{strings_strings}. (A single spatial configuration, like 
this one, of course corresponds to a particular realization of random
initial conditions.) 
This loop stays intact,
up to small vibrations, for a long time, being in quasi-equilibrium.
However, later on, different segments of the loop quickly approach
each other
and reconnect at $\tau =328$, forming another loop configuration
at $\tau =330$.
 This final loop collapses almost to a point,  bounces
once, collapses again and disappears.\footnote{The movie which shows this
process can be
seen at {\rm 
 http://www.physics.purdue.edu/$\sim$tkachev/movies.html}
or at {\rm http://physics.stanford.edu/$\sim$linde}. } 
This process of
reconnection and final collapse
confirms that loops are dynamical strings, not accidental lines in space.

To find the range of ${\rm v}$ for which strings are produced we had run
simulations with   different values of   ${\rm v}$  in the interval from
$4\times 10^{15}$ to $10^{17}$ GeV.  In all these cases we observed 
formation
of string loops. From cosmological perspective, however, the most 
interesting
possibility would be formation of an infinite string. 
One can expect that, at least when  ${\rm v}$ is large, the  probability 
of 
formation of long strings is higher
in the case when the moment of the symmetry breaking nearly coincides
with the moment when $\langle \phi_1 \rangle$ passes through  zero.
If that expectation is correct, the number of  long strings should be
a non-monotonic function of ${\rm v}$.

\begin{figure}[Fig9]
\centering
\leavevmode\epsfysize=8cm \epsfbox{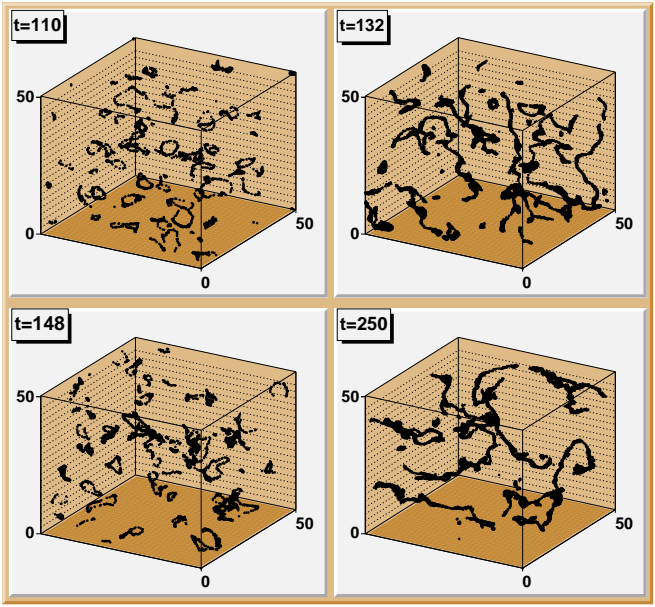}\\
\

\caption[Fig1]{\label{longstrings} The string distribution at the time $\Delta \tau = 10$ after the moment of the phase transition   for ${\rm v} =  10^{17}$ GeV, $6\times 10^{16}$ GeV, $5\times 10^{16}$ GeV, and $3\times 10^{16}$ GeV respectively. The plot for ${\rm v}=3\times 10^{16}$ GeV corresponds to a different
realization of the initial conditions than Fig. 5.}
\end{figure}

To verify this conjecture we plotted the string distribution at the time $\Delta \tau = 10$ after the  phase transition (which happens at different times) for ${\rm v} = 3\times 10^{16}$ GeV, $5\times 10^{16}$ GeV, $6\times 10^{16}$ GeV, and $10^{17}$ GeV,  for $\lambda = 10^{-13}$, see Fig. \ref{longstrings}. As we see, for   ${\rm v} = 5\times 10^{16}$ GeV  and $10^{17}$ GeV all loops are short, whereas for   ${\rm v} = 3\times 10^{16}$ GeV and $6\times 10^{16}$ GeV there are many large loops.  This indicates a possibility of formation of a network of  infinite strings for  ${\rm v} = 3\times 10^{16}$ GeV and $6\times 10^{16}$ GeV.

\begin{figure}[Fig9]
\centering
\leavevmode\epsfysize=7cm \epsfbox{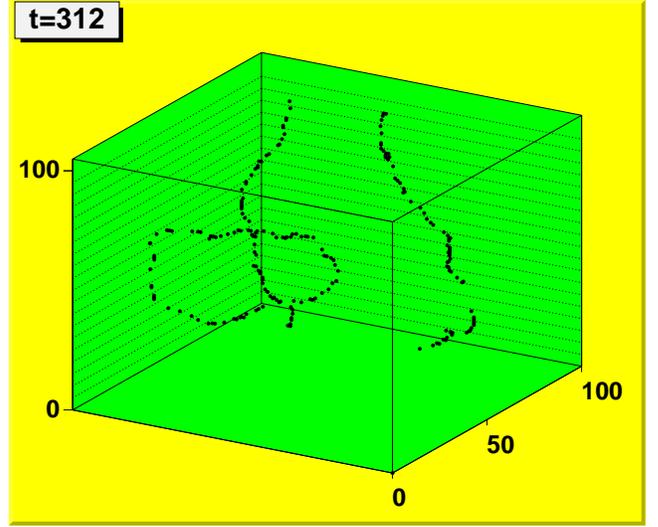}\\
\

\caption[Fig1]{\label{100} The string distribution     for ${\rm v} =  3 \times 10^{16}$ GeV in a box of a larger size.}
\end{figure}

Additional evidence for (or against) formation of infinite strings after
preheating may be obtained if one uses lattices of a greater  
size. In particular, we performed simulations for ${\rm v} = 3 \times 10^{16}$ GeV in a larger box ($L=32\pi$), see Fig. \ref{100}. At the time shown in
this figure, the physical size of the box is $L_{\rm phys}\sim (10^2) m^{-1}$, 
where $m^{-1} \sim 1/\sqrt\lambda {\rm v}$ is the typical thickness of the 
string. In Fig. \ref{100}, we see two ``infinite" strings and a large string 
loop. Note that on a lattice with periodic boundary conditions ``infinite" 
strings can only be created in pairs, because the winding number for our 
initial conditions is zero.

Also, one may consider models where one may have independent reasons to 
expect  
production of infinite strings. For
example, one may add to our model the term ${g^2\over 2}  \phi^2\chi^2$
describing interaction of the inflaton field $\phi$ with the scalar field
$\chi$ with the coupling constant $g^2 \gg \lambda$. We have studied
this model in Ref. \cite{kklt} for the case of a one-component field  
$\phi$
and found a first-order phase transition. We expect that this result  
remains valid for the two-component field as well, because the main reason
for the  first order phase transition found in \cite{kklt} was the strong
interaction of the field $\phi$ with the field $\chi$ rather than the
self-interaction of the field $\phi$.
If this is indeed the case, then nonthermal fluctuations of the field 
$\chi$
will lead to
symmetry restoration with respect to both of the fields $\phi_1$ and
$\phi_2$.
These two fields will be captured in the local minimum of the effective
potential at $\phi = 0$ until the moment of the first order phase
transition.
In this case, there will be strings formed by the usual Kibble mechanism,
including infinite ones. We
hope to return to the discussion of this model  in a separate
publication.

This work was supported in part by
DOE grant DE-FG02-91ER40681 (Task B) (S.K. and I.T.),
NSF grants PHY-9219345 (A.L.), PHY-9501458 (S.K. and I.T.),
 AST95-29-225 (L.K. and A.L.), and by the Alfred P. Sloan Foundation
(S.K.).

\

\

\end{document}